# Accuracy of simple, initials-based methods for author name disambiguation

*Staša Milojević*

*Indiana University, School of Informatics and Computing, Department of Information and Library Science*

smilojev@indiana.edu

## Abstract

There are a number of solutions that perform unsupervised name disambiguation based on the similarity of bibliographic records or common co-authorship patterns. Whether the use of these advanced methods, which are often difficult to implement, is warranted depends on whether the accuracy of the most basic disambiguation methods, which only use the author's last name and initials, is sufficient for a particular purpose. We derive realistic estimates for the accuracy of simple, initials-based methods using simulated bibliographic datasets in which the true identities of authors are known. Based on the simulations in five diverse disciplines we find that the *first initial* method already correctly identifies 97% of authors. An alternative simple method, which takes *all initials* into account, is typically two times less accurate, except in certain datasets that can be identified by applying a simple criterion. Finally, we introduce a new name-based method that combines the features of *first initial* and *all initials* methods by implicitly taking into account the last name frequency and the size of the dataset. This *hybrid* method reduces the fraction of incorrectly identified authors by 10-30% over the *first initial* method.

## 1 Introduction

For a significant fraction of studies that are based on bibliometric data, as well as for purposes of research evaluation, it is essential to be able to attribute specific bibliographic records to individual researchers. A practical problem with this straightforward step is that there is a certain level of ambiguity in this process, which is known as the *author name disambiguation* problem. The problem manifests itself in two ways: a given individual may be identified as two or more authors (*splitting*), or, two or more individuals may be identified as a single author (*merging*). Here we use the term *individual* to refer to an actual person, and an *author* to refer to an entity that results from the author disambiguation procedure. Both the splitting and the merging can happen as the result of the same disambiguation method.

The problem of author disambiguation fundamentally arises because personal names are not sufficiently distinct considering the large number of researchers active in most disciplines today. It is further exacerbated by the inconsistent way in which author names are reported in publications. The disambiguation method that is by far the most accurate is the manual inspection of full bibliographic records augmented with the examination of full-text articles and comparison with other sources of information, such as personal web sites, CVs, etc. Apparently, such a method is prohibitive for large-scale studies, which has led to the development of partially or completely *unsupervised* methods of author disambiguation. We divide unsupervised methods in two classes: *simple,* or name-based methods, which use only the information contained in authors' last name and initials, and *advanced* methods, which require additional information or exploit the relationships between author names.[1] We are aware of only two simple methods (e.g., (Newman, 2001)): the *first initial* method and the *all initials* method. The first initial method treats all authors with the same first initial as belonging to the same individual, while the all initials method treats authors with different middle initials but the same first initial as different entities. The latter method may appear more accurate, but its drawback is that it will treat an author with and without the middle initial as different individuals, while in fact it may be the same author whose middle initial is not consistently reported.

On the other hand, numerous advanced methods have been proposed, no doubt owing to the fact that the advanced name disambiguation represents an interesting problem in machine learning. A recent review by Ferreira, Gonçalves, and Laender (2012) lists 17 advanced methods. Comprehensive review of the problem and proposed solutions was also given by Smalhauser and Torvik (2009), while the probabilistic formalization of the problem can be found in Tang, Fong, Wang, and Zhang (2012). What is common to all advanced methods is that they use some measure of similarity to identify bibliographic records authored by the same individual. To achieve this goal they use various types of information in addition to the author name itself. This includes similarity of coauthor names, article titles, publication venues and subject headings (e.g., Torvik, Weeber, Swanson, and Smalheiser (2005), Cota, Ferreira, Nascimento, Gonçalves, and Laender (2010)), optionally supplemented with an external database of affiliations (D'Angelo, Giuffrida, & Abramo, 2011). Some recent studies also evaluate the knowledge base (references in articles) and citation counts, (L. Tang & Walsh, 2010), use the similarity of the full topology of collaboration networks (Amancio, Oliveira, & Costa, 2012), or include crowdsourced information (Sun, Kaur, Possamai, & Menczer, 2013).

A question that a researcher wishing to perform an author-centered bibliometric study faces is whether these advanced methods are worth the additional conceptual, programming, and computational effort. The rationale for introducing advanced methods is to improve the accuracy of simple methods, which are perceived as insufficient for specific purposes (Strotmann & Zhao, 2012) or problematic in general. Remarkably, no study to date provides the definitive answer to the simple question of whether the presumably high level of precision ensured by advanced methods is required for a large majority of studies in which one is interested only in general statistical trends, and which would not be affected by some small level of "contamination." Moody (2004) developed a rather sophisticated system of disambiguating authors based on the commonality of their first and last names and coauthorship patterns, but he found no significant difference in the results in coauthorship networks using the methods that he developed compared to relying only on an *all initials* approach. Furthermore, the current literature contains no assessments regarding which of the two simple methods (*first initial* or *all initials*) is more accurate.

---

[1] Some researchers consider only what we call advanced methods to represent genuine author disambiguation methods. However, the information contained in author last name and initials already forms the basis for author disambiguation at the basic level.

One possible reason why these issues have not received the necessary attention is that the determination of accuracy requires comparison with a large dataset of a priori perfectly disambiguated authors. In this study we show that such dataset can be produced using simulations that are based on the characteristics of the empirical data. One then subjects these simulated datasets to name-based disambiguation methods in order to determine their accuracy. In addition to establishing the accuracy of the existing simple methods, we propose a novel name-based method which combines the elements of the two traditional approaches, which we call the *hybrid* method. We show that this new simple method is often superior to either of the traditional name-based methods.

## 2   Methods

### 2.1   Features of the two simple author name disambiguation methods and the introduction of the hybrid method

Despite numerous advanced disambiguation methods that have been proposed in the literature, the simple methods, which are only based on names, are by far the most used in bibliometric studies. There are two such methods: the *first initial* method and the *all initials* method. The *first initial* method considers only the last name and the first initial. Any information contained in the middle initial is discarded. In this scheme names given as JACKSON, P and JACKSON, PA will be considered as belonging to the same individual. In this scheme, the name given as JACKSON, PS will also be attributed to the same author, which is apparently incorrect. The *first initial* method will therefore lead to some level of merging of distinct individuals into the same entity. The alternative method is to take into account *all initials*. Using this method JACKSON, P; JACKSON, PA and JACKSON, PS would all be considered distinct authors. This method may appear superior to the *first initial* method, but this is true only if the authors who have/use middle initial consistently report them on all of their publications. Unfortunately, one cannot assume that this is the case. A publication listing JACKSON, P may be authored by the same individual as either JACKSON, PA or JACKSON, PS. Consequently, the method using all initials can lead to substantial splitting of unique individuals into multiple authors.

Next, we show that another approach is possible, one that uses either the *first initial* or the *all initial* method depending on the case. Consider a large dataset containing thousands of bibliographic records. Let us say we have retrieved the name SMITH, C alongside SMITH, CM. Given that SMITH is a common last name and that our dataset is large, we feel that these are probably two different individuals, as would be judged by the *all initials* method. On the other hand, let us say that we have retrieved the names ZYWIETZ, C and ZYWIETZ, CM. Considering this name's rarity, we feel quite confident that this must be the same person, whose middle initial was not reported on some publications. In this case we would have preferred to use the *first initial* method that attributes these two names to the same author. It appears that a criterion can be introduced so that we use either the *all initials* or *first initial* method based on the frequency of a last name. However, this is not the case. Consider that our data set contained only 20 records (e.g., single-topic journal volume) and that we again encounter SMITH, CM and SMITH, C, perhaps first as the lead author and another time as a coauthor on multi-author article. Now it appears quite likely that these two names refer to the same person, but that the middle initial has not been reported consistently (presumably, the middle initial was omitted when this person was only a coauthor in a long list of authors). Apparently, the likelihood depends on the size of the dataset in addition to the last name frequency. Interestingly, there is a simple criterion to decide whether to split or merge two such names that does not involve any explicit assessment of name frequencies or dataset sizes. Instead, this criterion relies on information contained in the names themselves. Namely, in the case of a large dataset in which we found SMITH, C and SMITH, CM, we would probably also encounter a name SMITH, CJ (or some other middle initial). On the other hand, in the data set of only 20 articles SMITH, CM is most likely the only name with SMITH, C as the root. Thus, a criterion that implicitly takes into account both the intrinsic frequency of some last name and the size of the dataset is whether there appears more than *one* middle initial for a given name root (root is last name plus the first initial). If it does, then one would attribute all names to separate authors (i.e., use the *all initials* method). If there does not (i.e., there is only one middle initial for SMITH, C), then one would attribute the variants with and without the middle initial to the same author (i.e., use the *first initial* method). We call this simple method that uses either the *first initial* method or the *all initial* method depending on the context a *hybrid* method. Its schematic representation is given in Figure 1. Like the two traditional simple methods it requires no additional information beyond the names and is easy to implement.

Like other name-based methods, our newly proposed method cannot disambiguate names with the same first and middle initial that may belong to several individuals, which will more often affect Asian names. Instead, it makes a more appropriate choice between whether to take into account the information contained in the middle initial or not. Similarly, the new method is meant to be applied to author names alone, and it cannot be used to disambiguate other entities, such as institution names.

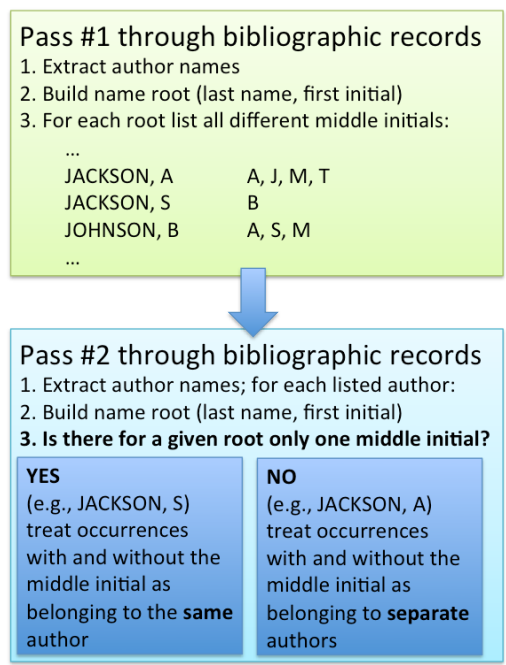

**Figure 1. Schematic representation of the proposed hybrid method of using initials to disambiguate author names.**

## 2.2 Evaluation and comparison of name-based author disambiguation methods

In order to evaluate the accuracy of a disambiguation method one needs to compare the results of the disambiguation with the true identity of authors. Perhaps the only effective way of accomplishing such evaluation for very large datasets is to produce a *simulated* dataset in which the true identity of each individual is known. For such evaluation to be realistic the corresponding simulated datasets must be built based on the measured characteristics of the empirical dataset.

### 2.2.1 Empirical datasets to be simulated

We chose articles from core journals in five disciplines published in a five year period between 2006 and 2010 as the empirical datasets to be simulated in order to test the performance of the two traditional simple methods and our *hybrid* method. We then selected five unrelated disciplines (astronomy, mathematics, robotics, ecology, and economics) in order to sample a range of dataset sizes, productivity patterns and name-reporting practices, all of which will affect the effectiveness of a disambiguation method. The list of core journals that were included in the dataset, as well as the rationale for their inclusion, had been given in Milojević (2012). In each discipline there were between four and nine journals. Bibliographic data for articles in these journals were obtained from Thomson Reuters' Web of Science. The number of articles in each of the five datasets is listed in Table 1. It ranges from a few thousand to over thirty thousand.

Table 1. Properties of empirical datasets. All data refer to 2006-2010 publication period.

| Field | Number of articles | Number of authors | Average productivity (in five years) | Percent of authors with middle initials | Reporting rate of the middle initial of authors who have it |
|---|---|---|---|---|---|
| Astronomy | 31,473 | 30,605 | 6.93 | 50% | 74% |
| Mathematics | 3,244 | 4,396 | 1.43 | 29% | 100% |
| Robotics | 2,630 | 5,734 | 1.54 | 31% | 76% |
| Ecology | 5,420 | 11,308 | 1.69 | 67% | 83% |
| Economics | 2,352 | 2,836 | 1.64 | 32% | 97% |

### 2.2.2 Simulation

We build the simulation to match the five empirical datasets. This is performed in the following way: We determine the number of authors in each dataset and the number of distinct *last* names. The former will obviously depend on the disambiguation method used, but such detail is irrelevant because it is sufficient for the simulation to be approximately the same size as the empirical dataset. Next, we assign a last name to each simulated author. This assignment is based on the frequency distribution of last names. To determine this distribution we use empirical data. One cannot simply construct the distribution of last names from the full, five-year dataset because most authors will have a different number of publications, which will lead to increased counts for some last names. In order to separate the effects of the productivity from the intrinsic name frequency we consider only the distribution of last names of first-listed authors in a single year. The limited time window and the focus on first authors results in most individuals appearing only once, thus ensuring that the increased frequency of some last names reflects the actual higher incidence of those names. We construct a frequency distribution using the astronomy dataset for each year separately (2006 through 2010) and then average the relative distributions. The resulting distribution is shown in Figure 2. The intrinsic frequency of last names is close to a power law with a slope $\alpha = -3.18$ (determined from a least square fit to logarithmically binned data points (Milojević, 2010)). This means that for a given common last name there are approximately 1,000 other last names that are 10 times less common. We assume that the frequency distributions of last names have the same power-law decline in all five fields.

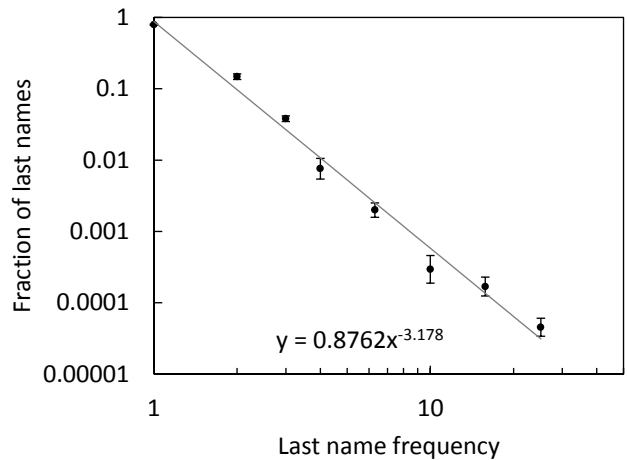

**Figure 2. Intrinsic frequency distribution of last names derived from astronomy dataset by combining data for each of the five years (2006-10). The distribution can be approximated as the power-law decline with a slope -3.18. This slope is used as the input for simulation. Scales are logarithmic. Error bars are standard deviations of the annual data points.**

Next, our simulation needs to take into account that a name of a given individual can appear on more than one publication, depending on the productivity in a given field. Productivity of each author is relevant in the context of this study because there will be more "opportunities" for inconsistent reporting of the middle initial in fields in which productivity is higher than in the fields in which each author appears on only a single paper. We use empirical data to determine the productivity distribution (regardless of author placement) in each of the five fields (Figure 3). The level of productivity varies significantly, from an average of 1.4 articles during five years in mathematics, to 6.9 in astronomy (Table 1). In all fields most authors have a single publication within the publication window. Based on these empirical distributions we assign to each simulated author a certain number of publications.

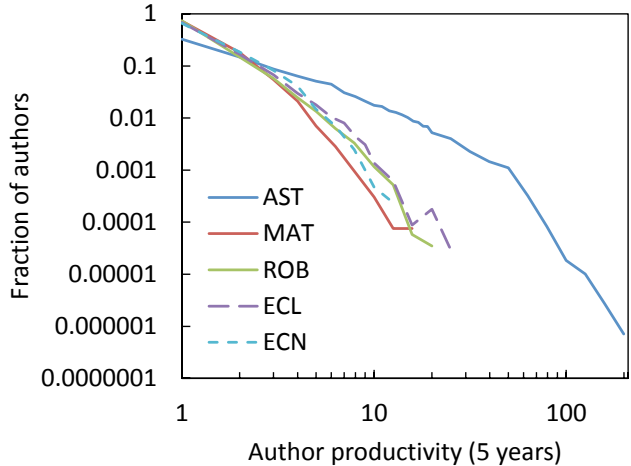

**Figure 3. Productivity distribution in the empirical datasets for 2006-2010. Productivity distribution in astronomy (AST) is less steep because authors typically have more articles due to longer coauthor lists. Distributions in other fields are more similar. Productivity of authors in simulations follows these exact distributions.**

We then assign first and middle initials, again following the frequency observed in empirical data. As an example, in Figure 4 we show the distribution frequencies of first and middle initials for astronomy. Letters A, J, and M are the most common and U and Q are the rarest. First and middle initials have similar but not identical distribution. Other disciplines have similar distributions. Not every author is assigned a middle initial. Its rate of occurrence is again made to match the empirical data. To determine empirically what fraction of authors have middle initials we need to keep in mind that even for authors who have and generally use middle initials, they will not always be reported (especially, for example if they are coauthors in large teams), so the overall occurrence rate will be lower. Determination of the reporting rate is impossible on an individual basis, which is why we perform a statistical determination based on some assumptions. We assume that the authors of *single-authored* articles will always report their middle initial. Therefore, we determine the *intrinsic* rate of middle initials from single-author papers alone. Then, in order to determine the reporting rate of the middle initial we compare the rate in single-author papers to that in all papers. For example, in astronomy 50% of single-author papers list a middle initial, while the frequency in the overall dataset (when these authors, are second, or third, or n-th coauthors) is 37%. Therefore, we take the intrinsic rate to be 50% and the reporting rate to be $0.37/0.50 = 73\%$. Intrinsic rate of middle initials varies from 29% in mathematics to 67% in ecology, while the reporting rate varies from 73% in astronomy to basically 100% in mathematics (Table 1). It appears that the usage of a middle initial is higher in fields with many researchers; perhaps because authors use them purposefully as a means of disambiguation. The reporting rate in fields with many coauthors per paper is lower, apparently because the corresponding authors on articles with many authors are not particularly diligent in reporting middle initials of their coauthors. Based on these empirically determined rates of middle initial usage we randomly choose which authors in the simulation will have a middle initial. Finally, based on the empirical reporting rate, we list this middle initial in some simulated publications of a given author and omit it in others. At each step in the construction of the simulated dataset we have verified that it correctly reproduces the empirical constraints.

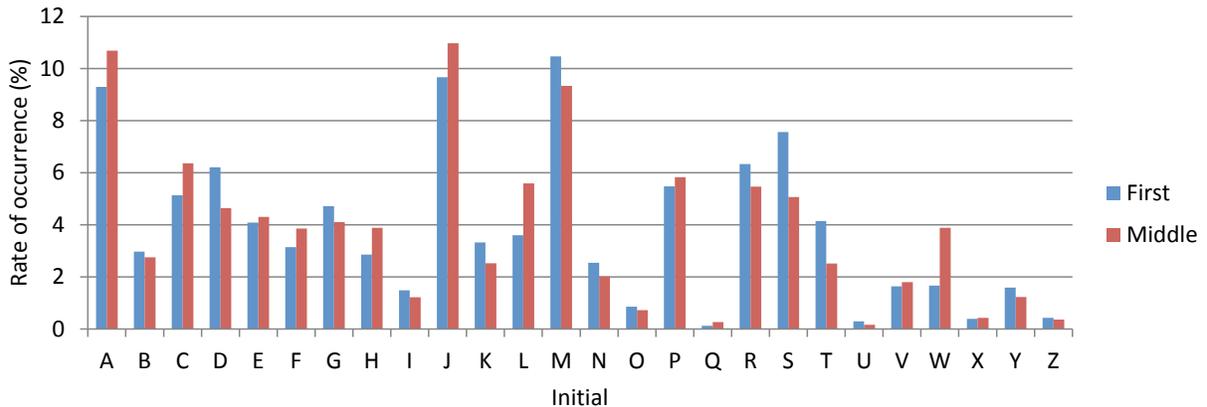

Figure 4. Distribution of occurrence of different initials. Data shown for the first initial (blue) and the middle initial (red) for astronomy (2006-10). Other fields have similar distributions. These distributions were used as inputs in the simulation alongside the rate of middle initial usage and the rate of its reporting, given in Table 1.

## 3   Results: Evaluation of disambiguation methods

The simulations now contain all of the information as the empirical datasets (author names from multiple publications, listed with or without middle initials), but with the crucial difference that the true identity of each distinct individual is known (each individual has a unique number in the simulation). As in the empirical dataset, some distinct authors in the simulation have the same names and some use and report middle initials while others do not. We now perform author name disambiguation in the simulated dataset in exactly the same way as we would in the empirical data. Each disambiguation method produces its own set of authors that it considers to be distinct. What metric is relevant to evaluate the accuracy of a given method? One could simply compare the number of disambiguated author names with the true number of authors. However, this is not a useful measure because the *all initials* method, as well as the *hybrid* method, can both split and merge authors making the end number not particularly relevant. Instead, we define the accuracy of each method as the percentage of true authors whose identity became compromised. This will include all individuals who were attributed to more than one author after the disambiguation, as well as all multiple real authors who became attributed to the same author.

The results of the evaluation are listed in Table 2, as the percentage of the authors whose identity is compromised. First, we see that a given method will perform differently in different datasets. The *first initial* method has relatively low "contamination" rates: between 2.5% and 6.1%. The *all initials* method span a wider range: from only 1.5% in the case of mathematics to very high "contamination" rate of 28.5% for astronomy. The *hybrid* method is more similar to *first initial* method with rates between 2.3% and 5.5%. Judging by the median percentage, the *hybrid* method outperforms both the *first initial* method (slightly, 2.8% vs. 3.1%) and the *all initials* method (2.8% vs. 5.6%). Of the traditional methods, the *first initial* method is typically better than the *all initials* method. However, in two of the five simulations the *all initials* method is actually more accurate than either of the two other methods. Both of these simulations (mathematics and economics) correspond to fields with a very high rate of reporting of the middle initial as well as small productivity. As we discussed previously, the *all initials* method is obviously superior if the middle initial is always used consistently, i.e. if the reporting rate was 100%. While we have used 100% and 97% for mathematics and economics, these estimates of the reporting rate could be somewhat higher than the actual ones. The additional reason why the *all initials* method outperforms the other two methods in mathematics and economics is that the authors in these fields have relatively low productivity, so for most of them a name will appear on only one publication, which implies that even non-perfect reporting rates of the middle initial are not detrimental. Indeed, had we assumed that astronomy, which has much higher productivity, had the same reporting rate as economics (97% vs. its actual 73%) the *all initials* method would still come out as the least successful (contamination rate of 10.6%) and the *hybrid* method as the best (4.7%). Similarly, had mathematics had only several percent lower reporting rate of the middle initial instead of the assumed perfect one, the *all initials* method would no longer be the most accurate, and the *hybrid* method would outperform it. We find that there is an empirical criterion that determines when the *all initials* method is preferred to the other two: the number of authors obtained using the *all initials* disambiguation must not be more than 2% larger than the number of authors obtained using the *first initial* method.

Table 2. Performance of different simple author disambiguation methods using simulated datasets. Results are given as the percentage of "actual" authors whose identity becomes compromised due to either splitting or merging in the process of disambiguation. Best method for a given dataset is in bold face.

| Simulated dataset corresponding to | First initial method | All initials method | Hybrid method |
|---|---|---|---|
| Astronomy | 6.1% | 28.5% | **5.5%** |
| Mathematics | 2.5% | **1.5%** | 2.3% |
| Robotics | 3.1% | 5.6% | **2.8%** |
| Ecology | 4.1% | 9.9% | **2.8%** |
| Economics | 2.6% | **2.2%** | 2.3% |

The results presented here reassure us that even without the application of advanced disambiguation methods the contamination levels in the *overall* datasets are low. The situation is likely to be different when the focus of a bibliometric study is on the extremes, such as determination of the most productive, most cited, or most collaborative authors. To assess how these basic methods would perform in such cases we produce lists of ~20 most productive authors in a simulated dataset corresponding to astronomy, as determined in each simple method, and compare them to the "true" list of most productive individuals. Using the *first initial* method the list of top 21 most productive authors contains all the authors from the true top 21 list.[2] The only difference is that one author (who in reality is ranked fourth) was merged with another, much lower ranked author which has boosted his/her productivity to the top spot. Ranking based on the *hybrid* method is the same. On the other hand, the top 22 list based on the *all initials* method contains only 11 authors who are actually ranked in the top 22. This emphasizes that the *all initials* method should not be used except in the special circumstances explained previously.

---

[2] Number is not exactly 20 because of a tie.

# 4 Conclusions and recommendations for use

We have shown that the traditional simple methods of author disambiguation within a single discipline or a field that use only information contained in names are quite accurate, with the percentage of authors whose identity becomes compromised (the "contamination" rate) typically being only several percent. Of the two traditional methods the *first initial* method is usually more accurate than the *all initials* method, but the exact performance will depend on the characteristics of the dataset. We have found a simple criterion to determine when this is the case. If the number of authors obtained using the *all initials* disambiguation is no more than 2% greater than the number of authors obtained using the *first initial* disambiguation, then the *all initials* method is the best. Our newly introduced *hybrid* method brings about 10-30% improvement over the *first initial* method in the five cases that we studied, and is, in general cases, the most preferred of the three methods. Its typical contamination rate is only 3%.

Remarkably, the accuracy of these simple methods, especially the *hybrid* method and *first initial* method (~97%) is not dramatically lower than that of very complex disambiguation schemes (~99%, (Torvik & Smalheiser, 2009)). Thus, the application of just the basic disambiguation methods will not have an adverse effect on many or most statistical bibliometric studies. We show that even when the focus is on the extremes ("most cited", "most productive", "best connected") the accuracy of two of the simple methods (*first initials* and hybrid) is very high (top 20 most productive authors are all correctly identified). Note that studies can verify whether their results are sensitive to the name disambiguation problem by simply performing a parallel analysis that excludes up to several percent of the most common last names (where last name frequency has been determined in a way that it is not affected by author productivity, see Section 2.2). Guided by the results presented in this paper we believe that they will in most cases confirm the robustness of the results obtained with only the simple disambiguation methods.

## Acknowledgments

I thank two anonymous referees for useful suggestions, and Colleen Martin for copy editing.